\documentclass{aa}  
\usepackage{graphicx}
\usepackage{upgreek}
\usepackage{txfonts}
\usepackage{bm}
\usepackage{xcolor}

\begin{document}

\title{On the degree of stochastic asymmetry in the tidal tails of star clusters}

\author{
  J. Pflamm-Altenburg
  \inst{1}
  \and
  P. Kroupa\inst{1,2}
    \and
  I. Thies
  \inst{1}
  \and
  Tereza Jerabkova\inst{3}
  \and
  Giacomo Beccari \inst{3}
  \and
  Timo Prusti \inst{4}
  \and
  Henri M. J. Boffin \inst{3}
}

\institute{Helmholtz-Institut f\"ur Strahlen- und Kernphysik (HISKP),
  Universit\"at Bonn, Nussallee 14–16, D-53115 Bonn, Germany\\
  \email{jpa@hiskp.uni-bonn.de}
  \and
  Charles University in Prague,
  Faculty of Mathematics and Physics, Astronomical Institute,
  V Hole\v{s}ovi\v{c}k\'{a}ch 2, CZ-180 00 Praha 8, Czech Republic
  \and
  European Southern Observatory, Karl-Schwarzschild-Strasse 2, D-85748 Garching bei M\"unchen, Germany
  \and
  European Space Research and Technology Centre (ESA ESTEC), Keplerlaan 1, 2201 AZ Nordwijk, Netherlands
}

\date{Received \ldots; accepted \ldots}

% \abstract{}{}{}{}{} 
% 5 {} token are mandatory

\abstract
    {Tidal tails of star clusters are commonly understood to be
      populated symmetrically. Recently, the analysis
      of Gaia data revealed large asymmetries between the leading and
      trailing tidal tail arms of the four
      open star clusters Hyades, Praesepe, Coma Berenices and NGC~752.}
    {As the evaporation of stars from star clusters into the tidal
      tails is a stochastic process,
      the degree of stochastic asymmetry is quantified in this work.}
    {For each star cluster 1000~configurations of test particles are
      integrated in the combined potential of
      a Plummer sphere and the Galactic tidal field over the life
      time of the particular star cluster.
      For each of the four star clusters the distribution function of the
      stochastic asymmetry is
      determined and compared with the observed asymmetry.}
    {The probabilities for a stochastic origin of the observed asymmetry of the
      four star clusters are: Praesepe $\approx$1.7~$\sigma$,
      Coma Berenices $\approx$2.4~$\sigma$, Hyades $\approx$6.7~$\sigma$,
      NGC~752 $\approx$1.6~$\sigma$.}
    {In the case of Praesepe, Coma Berenices and NGC~752
      the observed asymmetry can be interpreted as a stochastic evaporation event.
      However, for the formation of the asymmetric tidal tails of the Hyades
      additional dynamical
      processes beyond a pure statistical evaporation effect are required.}
    
    \keywords{
      open clusters and associations: general - 
      open clusters and associations: individual: Hyades, Praesepe, Coma Berenices, NGC~752 -
      stars: kinematics and dynamics
    }
    
    \maketitle
    %
%-------------------------------------------------------------------

    \section{Introduction}
    In general, stars form spatially confined in the densest regions
    of molecular clouds \citep{lada2003a,allen2007a}.
    After their formation different
    processes lead to the loss of stellar members:
    
    \emph{Early gas expulsion}: 
    The gas in the central part of the newly formed star cluster
    has not been completely converted into stars. Once the most massive stars
    have ignited the ionising radiation heats up the remaining gas
    leading to its removal from the star cluster. The initially virialised
    mixture of stars and gas turns into a dynamically hot and expanding
    star cluster. After loosing a substantial amount of members the remaining
    star cluster revirialises now having  a larger diameter than at its birth.
    This process only takes a few crossing times which are typically of the
    order of a few Myr for open star clusters \citep[e.g.][]{baumgardt2007a}.

    \emph{Stellar ejections}:
    In the Galactic field OB-stars are observed moving with much higher
    velocities than the velocity dispersion of the young stellar
    component in the Galactic field. They are assumed
    to be ejected form young star clusters either by the disintegration
    of massive binaries where one component explodes in a supernova
    (\emph{supernova ejection}) leaving the second component with its high
    orbital velocity,
    or by close dynamical interactions of multiple stellar
    systems (e.g. in binary-binary encounters) with energy transfer between the
    components. By orbital shrinkage of one binary
    potential energy is transferred to the other binary leading to its
    disintegration and leaving the two components with high kinetic
    energy behind
    \citep[e.g.][]{poveda1967a,pflamm-altenburg2006a,oh2016a}.
    In both cases, the velocity of the star is higher than the
    escape velocity of the star cluster allowing the particular star to escape
    from the star cluster into the Galactic field.

    \emph{Stellar evolution}:
    Due to stellar evolution, the total mass of the star cluster decreases
    continuously, reducing the binding energy and the
    tidal radius of the star cluster. The negative binding energy of stars
    which are only slightly bound may turn to a positive value.

    \emph{Evaporation and tidal loss of stars}:
    Gravitationally bound stellar systems embedded
    in an external gravitational field lose members due to
    the tidal forces. In the frame of a
    star cluster its potential well is lowered by the tidal forces
    at two opposite points known as Lagrange points. The velocities of
    stars in the Maxwell tail of the velocity distribution are sufficiently
    high to escape through these two Lagrange points. Each of both streams of
    escaping stars forms an elongated structure pointing in
    opposite directions, these being the tidal tails. If the size of the
    star cluster is small enough compared to the spatial change of the external
    force field then both depressions of the star cluster potential at the
    Lagrange points are equal.
    This is typically the case for star clusters in the solar
    vicinity orbiting
    around the Galactic centre.
    Thus, both streams of stars through the Lagrange points are equal and the
    tidal pattern is expected to be symmetric with respect to the star
    cluster \citep[e.g.][]{kuepper2010a}. Therefore,
    both tidal arms should be equal.

    Recently, the detailed analysis of Gaia data revealed asymmetries in the
    tidal tails of the four open star clusters Hyades \citep{jerabkova2021a},
    Preasepe, Coma Berenices \citep{jerabkova2022a}
    and NGC~752 \citep{boffin2022a}, challenging the common assumption
    of symmetric tidal tails.
    However, as it can not be predicted
    through which Lagrange point a star will
    escape from the star cluster into one of the tidal arms, evaporation
    can be interpreted as a stochastic process. Thus, different 
    numbers of members in both tidal tails of an individual star cluster
    are expected. It is the aim of this work to quantify the degree
    of asymmetry in tidal tails due to the stochastic population
    of tidal tails and to compare with the observations.

    In Sect.~\ref{sec_meassure} we describe how (a)symmetry in
    tidal tails of star clusters is quantified
    throughout this work.
    The data of the observed star clusters and the derived
    asymmetries are presented in Sect. \ref{sec_observations}.
    The evaporation process of the numerical Monte Carlo model is
    outlined in Sect.~\ref{sec_monte_carlo} including
    the determination of statistical asymmetries due to stochastic evaporation.
    Section~\ref{sec_theo} compares the numerical asymmetries
    with a theoretical model which interprets the
    evaporation of stars through both Lagrange points into
    the tidal tails as a Bernoulli experiment.
    
    \section{Measuring the (a)symmetry of tidal tails}\label{sec_meassure}
    In order to explore the degree of (a)symmetry in the tidal tails
    the distance distribution of tail members is quantified.
    \citet{kuepper2010a} performed $N$-body simulations and calculated
    the stellar number density as a function of the distance to the
    star cluster along the star cluster orbit. In their simulations 21
    models covering a range of different initial conditions such
    as the galactocentic
    orbital radius or the inclination of the orbit
    were computed with initially 65536 particles using the N{\sc nody4} code
    \citep{aarseth1999a,aarseth2003a}
    which performs a full force summation over all particles.
    This high number of particles
    leads to a smooth population of the tidal arms. Thus, only small statistical
    fluctuations are expected. In order to explore the statistical fluctuations
    of the member number of
    tidal tails of open star clusters with only a few hundred stars
    in the tidal tails
    multiple simulations of star clusters resembling the observed ones are
    required.
    
\begin{figure}
     \centering
     \includegraphics[width=\columnwidth]{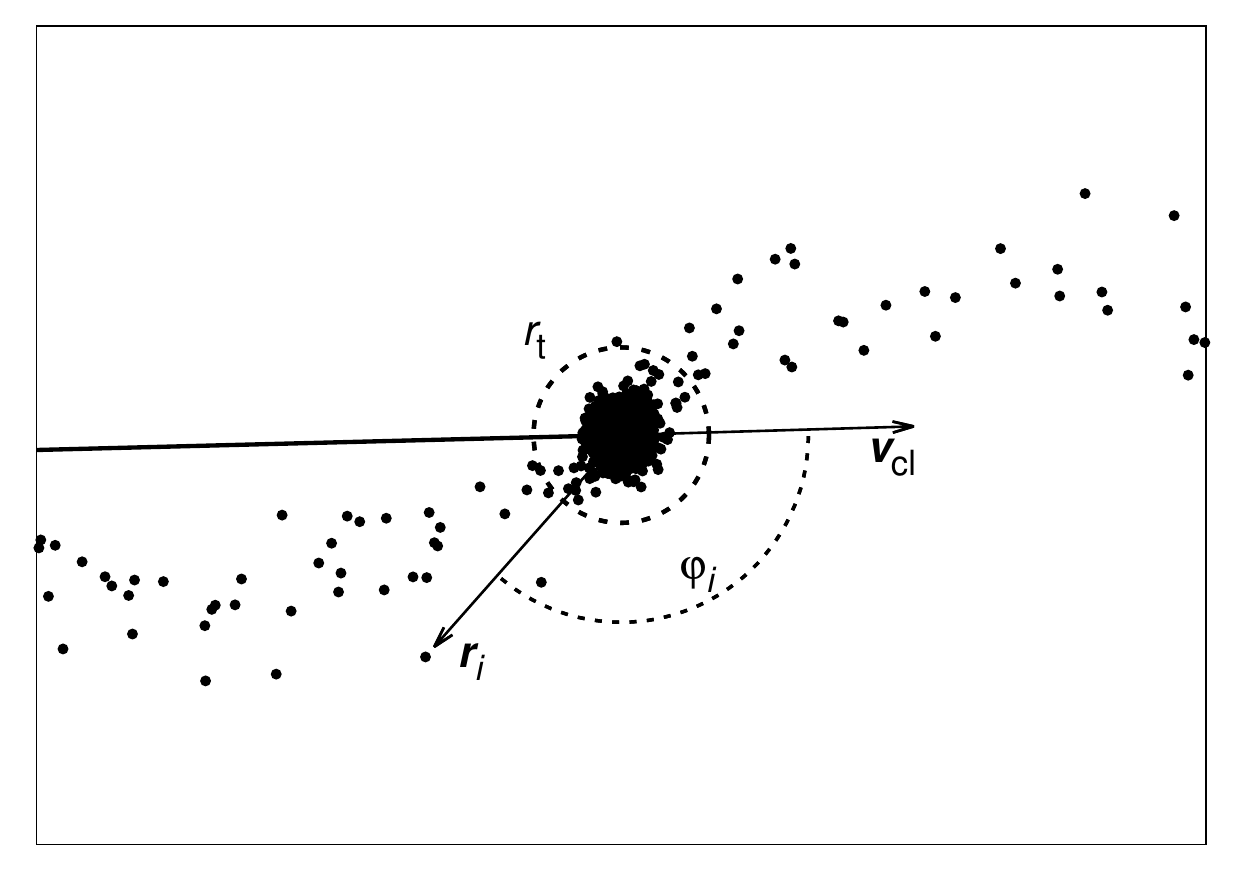}
     \caption{Sketch of the method to quantify the (a)symmetry in tidal tails.
       The angle of the orientation, $\varphi_i$, of each star is the angle
       enclosed by the position vector, $\vec{r}_i$, of the particular star
       with respect to the centre of the star cluster
       and the actual velocity vector, $\vec{v}_\mathrm{cl}$, of the star
     cluster.}
     \label{fig_measure_assymetry}
   \end{figure}

    In order to avoid effects by uncertainties of the Galactic tidal field,
    here the actual velocity vector of the star cluster is used as the reference
    (Fig.\ref{fig_measure_assymetry}) and not the distance to the cluster
    along the orbit of the star cluster.
    The membership and distance of each star is
    determined as follows: For each star its distance is given by
    its position vector, $r_i = |\vec{r}_i|$,
    in the star cluster reference frame. If its distance is larger than the
    tidal radius, $r_\mathrm{t}$, it is a tidal tail star.
    The orientation angle, $\varphi_i$,
    is the angle enclosed by the position vector, $\vec{r}_i$,
    of the star and the actual
    velocity vector of the star cluster, $\vec{v}_\mathrm{cl}$. 
    A star is considered to be a member of the leading arm if
    $0 \le \varphi_i \le 90^\circ$, 
    and to be a member of the trailing arm if
    $\varphi_i > 90^\circ$. Here, the membership of a star
    to belong to either the leading or the trailing arm is a sharp criterium
    without any probability. For those stars located very close to the border
    separating the leading from the trailing arm the errors of the
    observed data of the positions of the stars are not considered.
    
    The normalised asymmetry, $\epsilon$,
    is given by the difference of the number of members in the leading
    tail, $n_\mathrm{l}$, and the number of members in the trailing
    tail, $n_\mathrm{t}$, divided by the total number of tidal tail stars, $n$,
    \begin{equation}\label{eq_norm_asymmetry}
      \epsilon = \frac{n_\mathrm{l}-n_\mathrm{t}}{n_\mathrm{l}+n_\mathrm{t}}=\
     \frac{n_\mathrm{l}-n_\mathrm{t}}{n}\,.
    \end{equation}

   \section{Observed open star clusters}\label{sec_observations}
   In this section the structure of the tidal tails of the Hyades, Praesepe,
   Coma Berenices and NGC~752 are analysed using the method described in
   Sect.~\ref{sec_meassure}. The present-day stellar distributions
   have been obtained by the \citet{jerabkova2021a} compact convergent
   point (CCP) method which allows the tidal tails to be mapped
   to their tips.

   Table~\ref{tab_sc_data} summarises the observed values of the star clusters which are of interest for
   this work (see \citealt{kroupa2022a} for details).
     
   \begin{table}
     \begin{tabular}{ccccc}
       star cluster      & Hyades  & \parbox[c]{1cm}{Coma\\Berenices} & Praesepe  & NGC~752\\
       \hline
       $d$ / pc  & 47.5 & 85.9 & 186.2 & 438.4\\ 
       $M$ / M$_\sun$      &  275       &  112      &  311        & 379\\
       $b_\mathrm{Pl}$ / pc &  3.1       &  2.7      &   3.7        & 4.1\\
       $r_\mathrm{t}$  / pc &  9.0       &  6.9      & 10.8         & 9.4 (1.2285$^\circ$)\\
       $t$ / Myr           &  680       & 750       &  770         & 1750\\
       $n$                 & 541        & 640       & 833          & 298\\
       $n_\mathrm{l}$       & 351        & 348       & 384          & 163\\
       $n_\mathrm{t}$       & 190        & 292        & 449         & 135\\
       $\epsilon$          &  0.298     & 0.088      &-0.078       & 0.094\\
       $n_\mathrm{l,50-200}$&  162       & 133        & 87          & 56\\
       $n_\mathrm{t,50-200}$&  64        & 111        & 140         & 43\\
        \hline
       $x$ / pc            &   -8344.61 & -8305.78   & -8439.78    & -8294.05\\
       $y$ / pc            &   -0.23    &  -5.84682  & -69.384     &  275.07\\
       $z$ / pc            &   10.69    &   112.516  &  128.552    & -158.408 \\
       $v_x$ / km s$^{-1}$  &  -32.01    &   8.69065  &  -33.3617   & -8.19\\
       $v_y$ / km s$^{-1}$  &  212.37    & 226.515    &  210.711    & 216.262\\
       $v_z$ / km s$^{-1}$  &   6.44     &   6.16155  &   -1.83814  & -12.81\\       
     \end{tabular}
     \caption{\label{tab_sc_data}
       Data of the star clusters. ({\it upper section}:)
       $d$: distance Sun--star cluster,
       $M$: current stellar mass within tidal
       radius $r_\mathrm{t}$, $b_\mathrm{Pl}$:
       Plummer parameter obtained from the observed half mass
       radius, $r_\mathrm{t}$:
       observed tidal radius, $t$: average of published ages of the
       star clusters,
       $n$: total number of tidal tail members,
       $n_\mathrm{l}$: number of members in the leading tidal tail,
       $n_\mathrm{t}$: number of members in the trailing tidal tail,
       $n_\mathrm{l,50-200}$: number of members in the leading tidal tail at a distance
       of 50--200\,pc from the cluster centre,
       $n_\mathrm{t,50-200}$: number of members in the trailing tidal tail at a distance
       of 50--200\,pc from the cluster centre.
       ({\it lower section}:) current position and velocity of
       the star clusters in
       the Galactic inertial rest frame used for the Monte Carlo simulations
       for an assumed Solar distance of 8.3 kpc to the Galactic centre, and
       27\,pc  above the Galactic plane  and a local rotational velocity
       of 225\,km/s as in \citet{jerabkova2021a}. The velocity components
       of the Sun
       in the Galactic rest frame are 11.1\,km/s towards to the Galactic centre,
       232.24\,km/s into the direction of Galactic rotation
       and 7.25\,km/s in positive vertical direction.
       Thus, the peculiar velocity of the Sun is [11.1, 7.24, 7.25]\,km/s.
       See \citet{kroupa2022a} for details.
     }
   \end{table}

\begin{figure}
     \centering
     \includegraphics[width=\columnwidth]{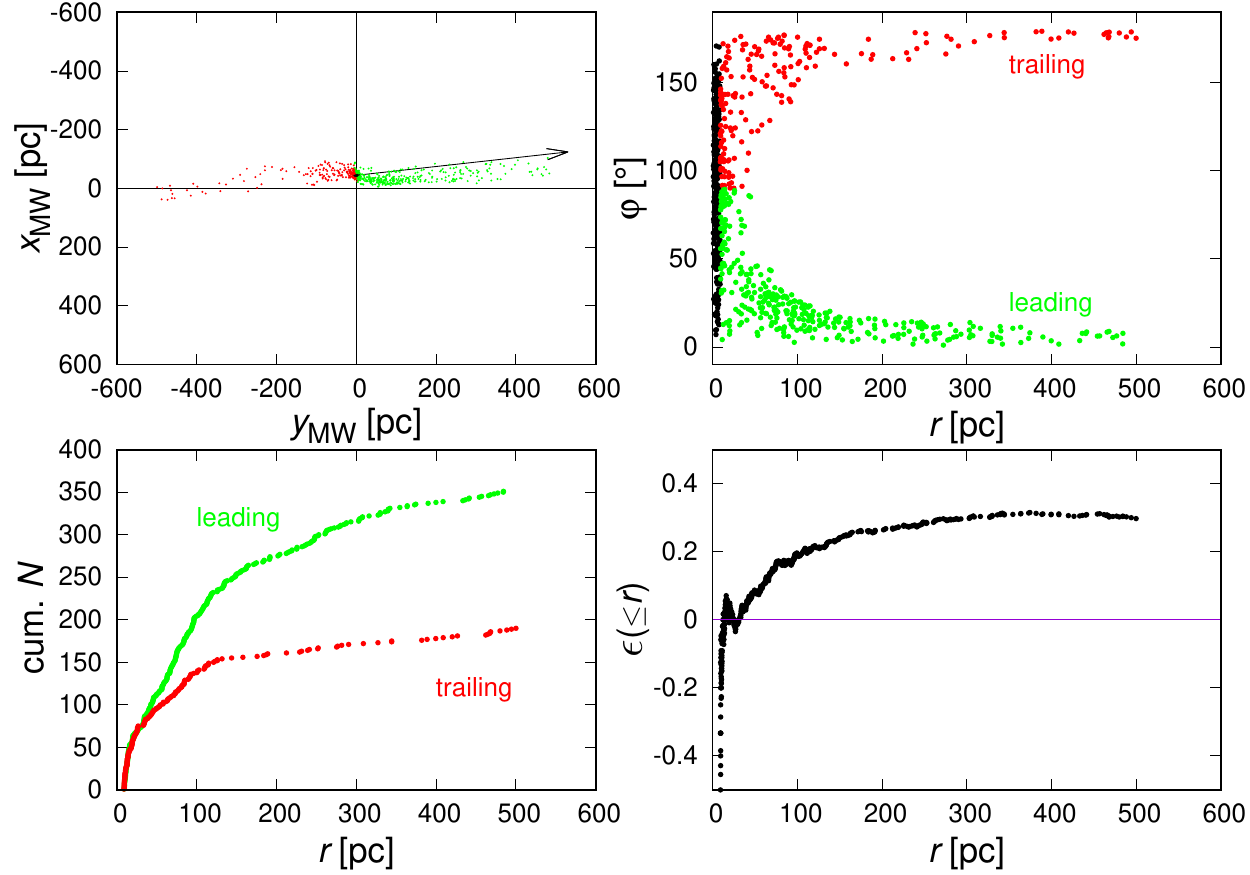}
     \caption{Hyades: (\emph{upper left:}) Shown is the spatial
       distribution of the members of the Hyades (black),
       the leading tidal arm (green) and the trailing tidal arm (red)
       from \citet{jerabkova2021a}.
       The coordinates
       $x_\mathrm{MW}$ and $y_\mathrm{MW}$ refer to a non-rotating rest frame
       of the Milky Way with a shifted origin where the position of the Sun
       lies at (0,0).
       The positive $x$-axis points towards the Galactic centre, the positive
       $y$-axis towards Galactic rotation. 
       The arrow indicates the motion of the star
       cluster centre in the Galactic rest frame. (\emph{upper right:})
       Shown are the
       positions of cluster and tidal tidal members in polar coordinates
       calculated
       as described in Sect.~\ref{sec_meassure}.
       (\emph{lower left:}) Radial cumulative 
       distribution of the number of 
       stars in the leading and trailing tidal arm of the Hyades 
       (\emph{lower right:}) Radial cumulative evolution of
       the normalised asymmetry (Eq.~\ref{eq_norm_asymmetry})
       considering tidal tail stars with a distance to the cluster
       centre smaller or equal to $r$.}
     \label{fig_hyades}
   \end{figure}

\begin{figure}
     \centering
     \includegraphics[width=\columnwidth]{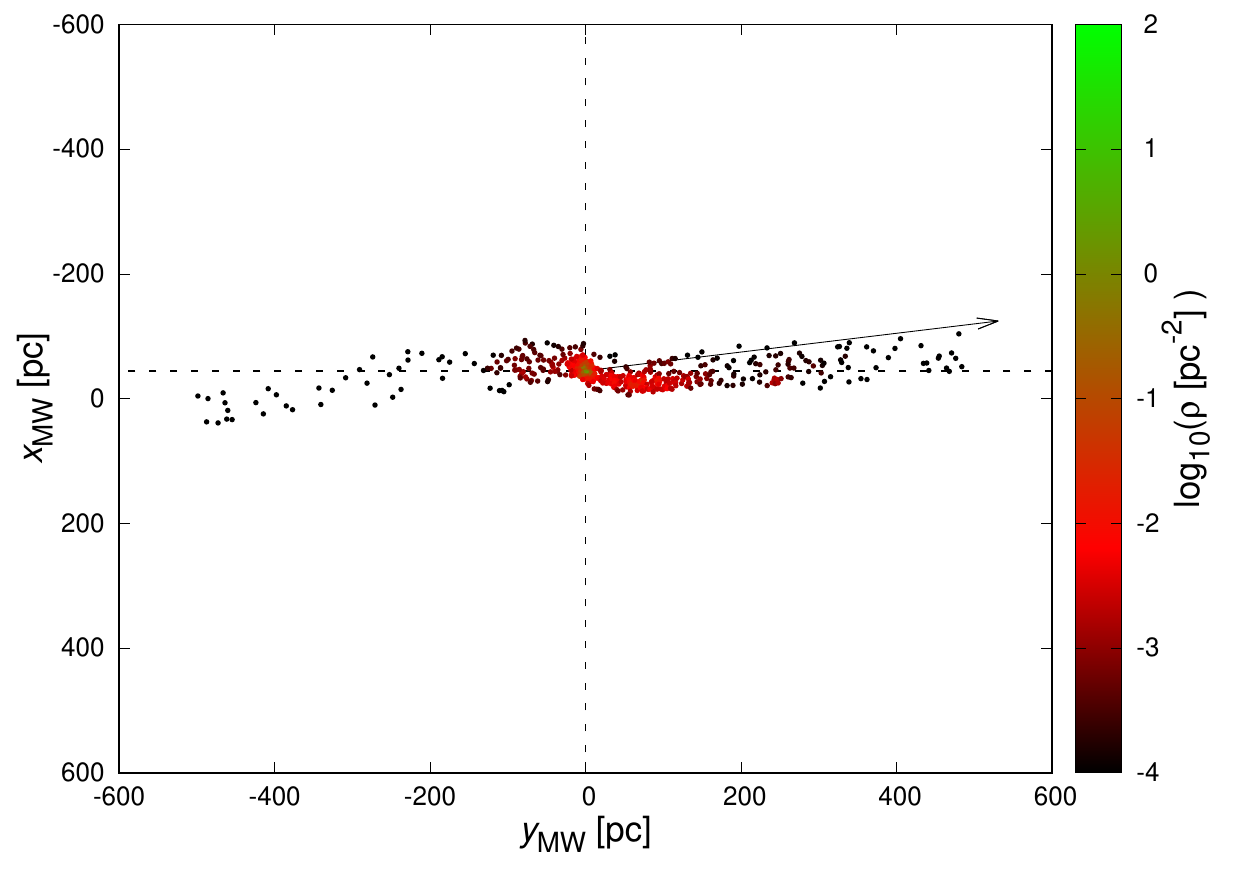}
     \caption{Surface number density of stars of the Hyades
       (cluster plus both tidal arms). Each dot represents one star.
       The local number density is calculated with the $6$th-nearest neighbour
       method \citep{casertano1985a} and colour-coded.
       The black arrow indicates the velocity vector
       of the Hyades star cluster in the Galactic rest frame. The centre of the
       star cluster is located at the intersection of the two thin
       dashed lines.}
     \label{fig_hyades_density}
   \end{figure}

\begin{figure}
     \centering
     \includegraphics[width=\columnwidth]{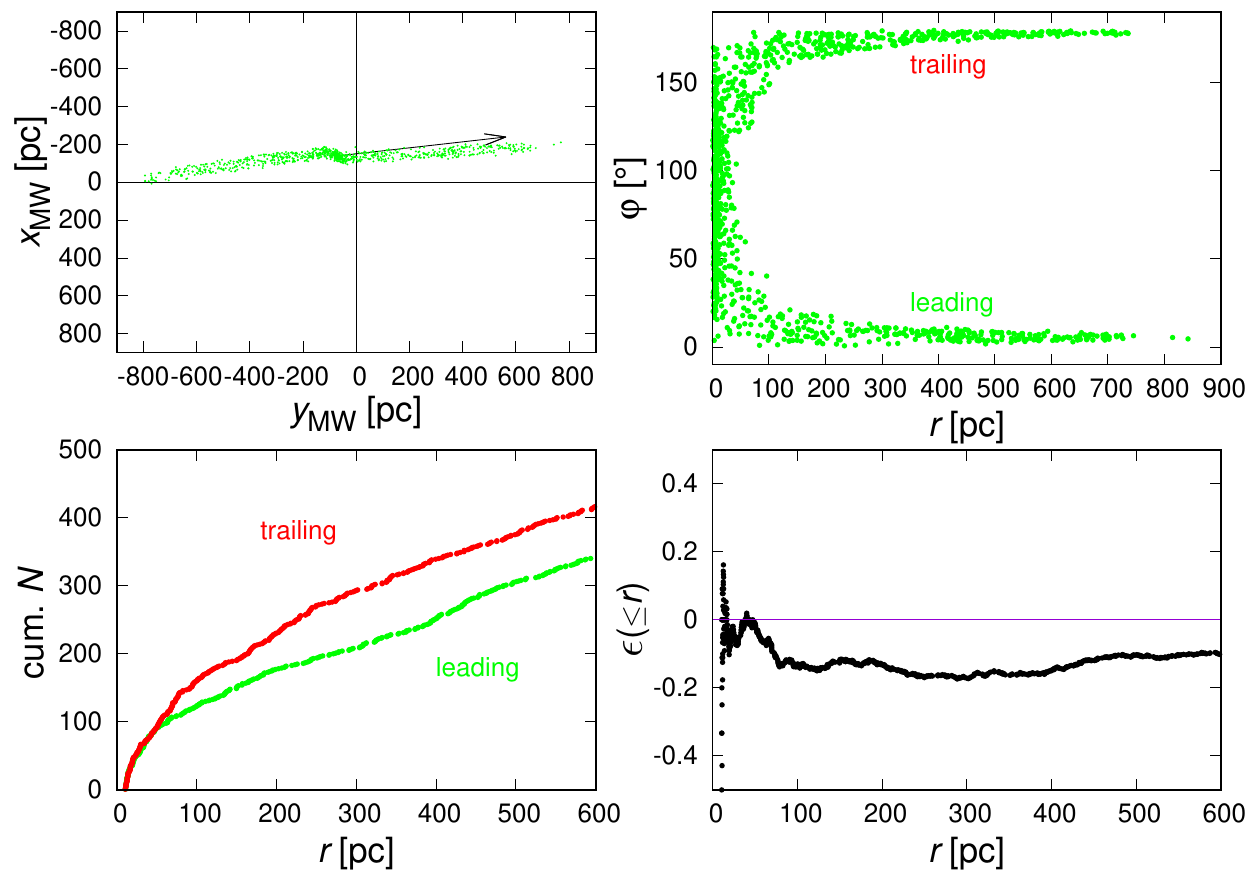}
     \caption{Praesepe from \citet{jerabkova2022a}: Similar to  Fig.~\ref{fig_hyades}.}
     \label{fig_praesepe}
   \end{figure}

\begin{figure}
     \centering
     \includegraphics[width=\columnwidth]{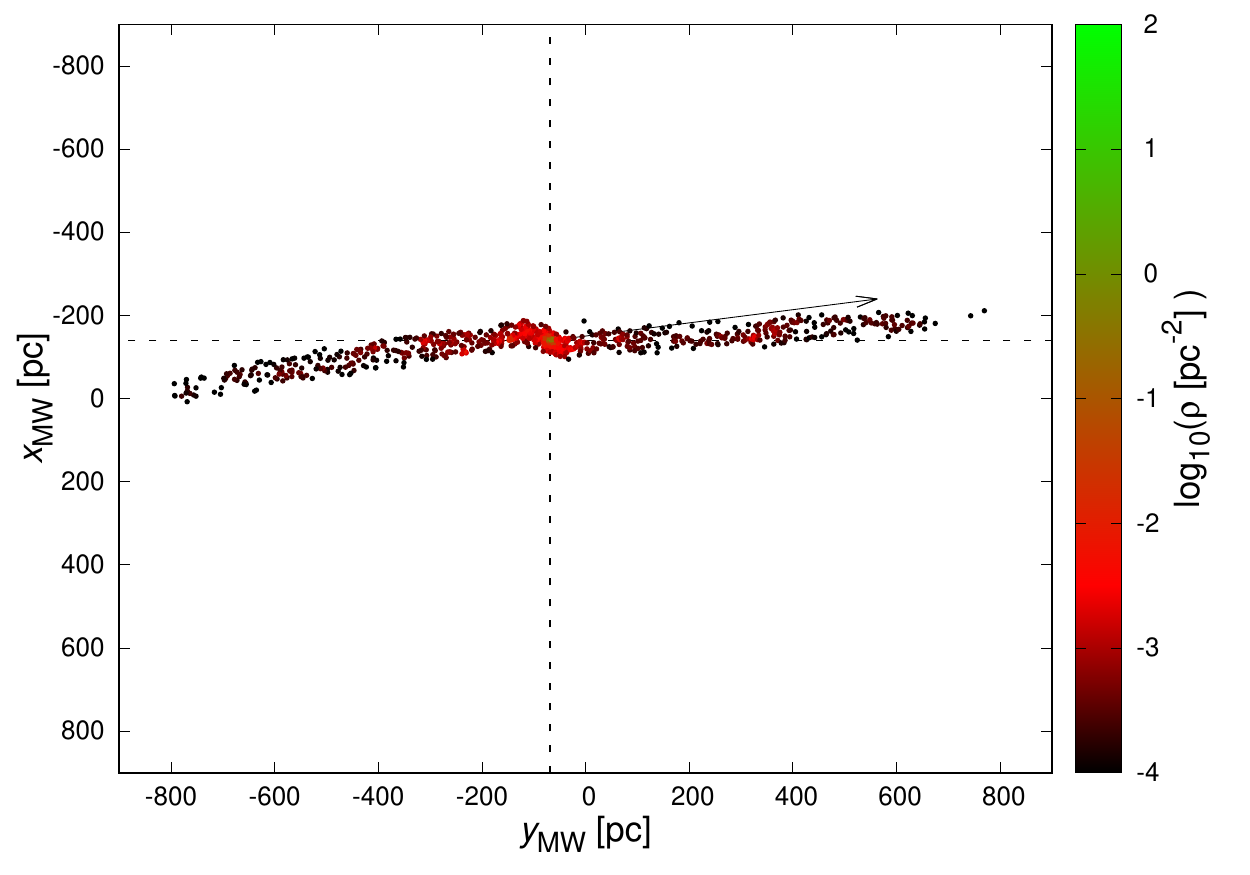}
     \caption{Praesepe: Similar to  Fig.~\ref{fig_hyades_density}.}
     \label{fig_praesepe_density}
   \end{figure}

\begin{figure}
     \centering
     \includegraphics[width=\columnwidth]{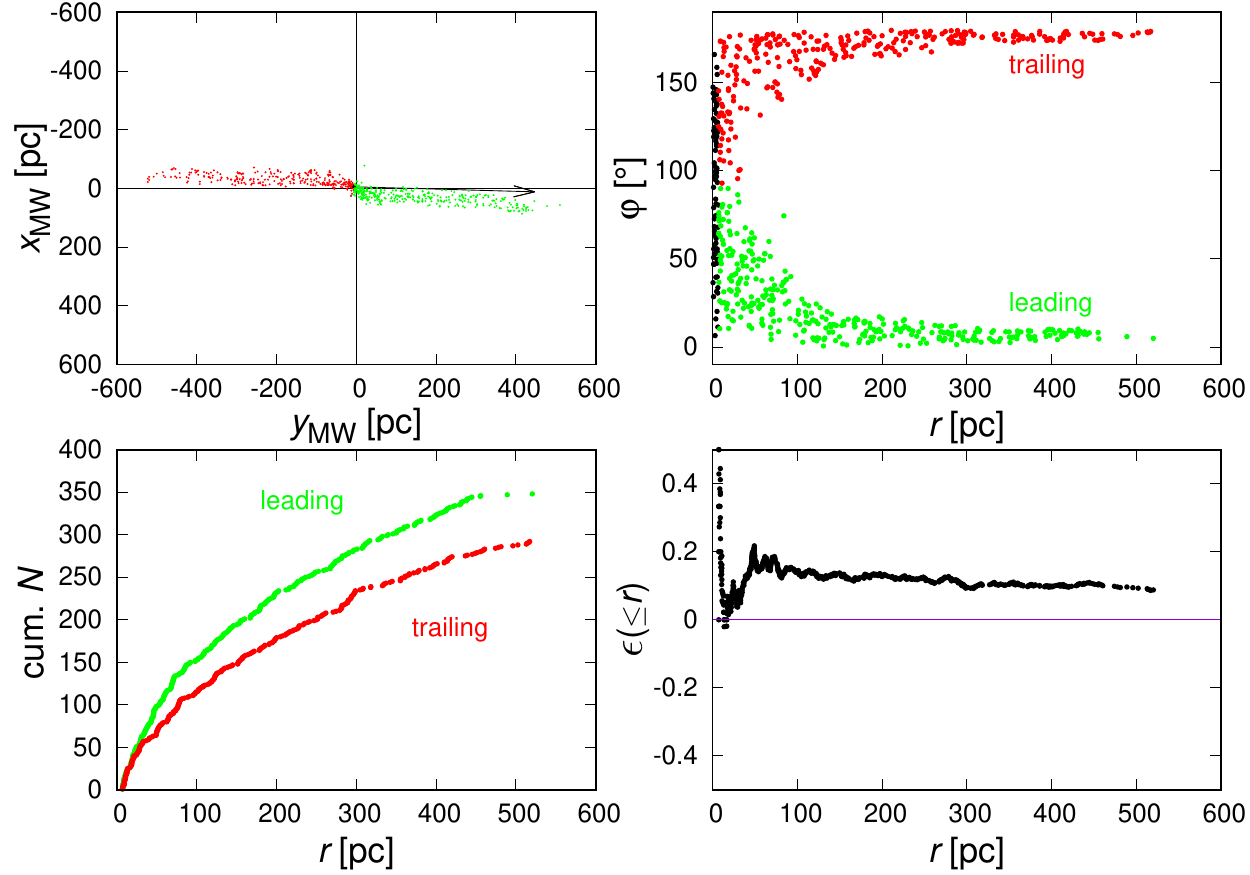}
     \caption{Coma Berenices from \citet{jerabkova2022a}:  Similar to Fig.~\ref{fig_hyades}.}
     \label{fig_coma_berenices}
   \end{figure}

\begin{figure}
     \centering
     \includegraphics[width=\columnwidth]{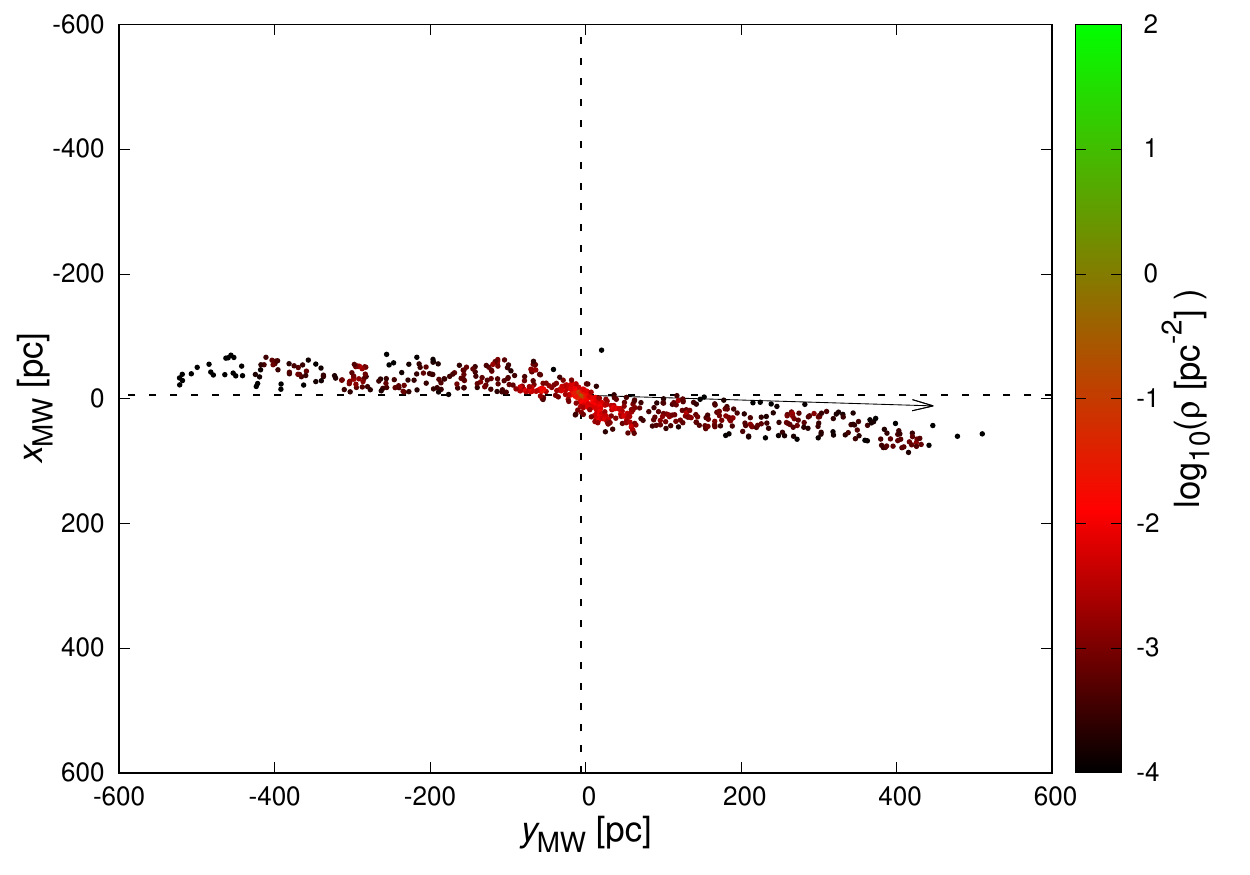}
     \caption{Coma Berenices: Similar to  Fig.~\ref{fig_hyades_density}.}
     \label{fig_coma_berenices_density}
   \end{figure}

\begin{figure}
     \centering
     \includegraphics[width=\columnwidth]{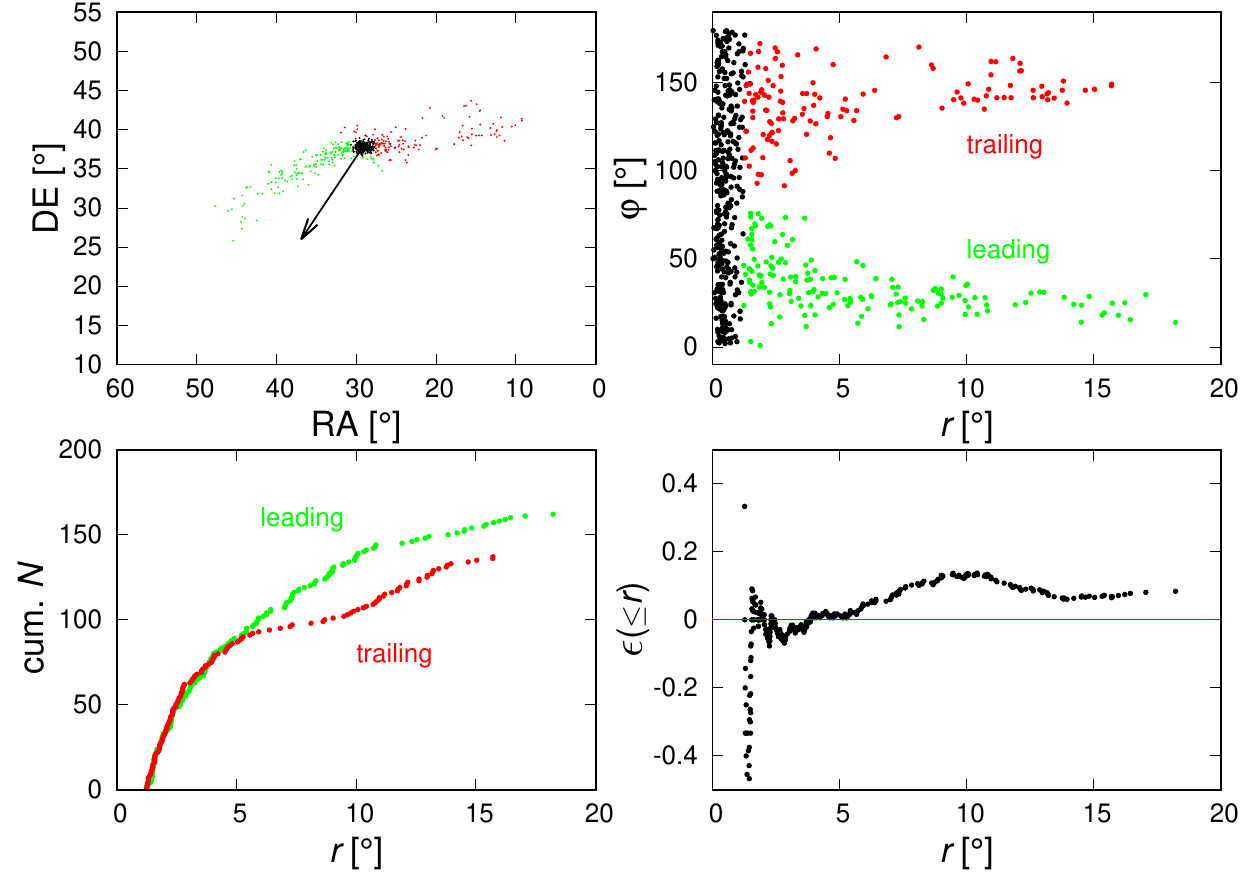}
     \caption{NGC~752 from \citet{beccari2022a}:  Similar to Fig.~\ref{fig_hyades}.}
     \label{fig_ngc752}
   \end{figure}

   The results of the analysis of the observational data are shown
   in Figs.~\ref{fig_hyades}--\ref{fig_ngc752}.
   The data for the Hyades are published in \citet{jerabkova2021a}
   and the detailed analysis of its tidal tails can be seen in
   Fig.~\ref{fig_hyades}. The upper left panel shows the spatial distribution of
   all stars associated with the Hyades. These are all stars, which are
   identified to be cluster members, that is, have a distance to the cluster centre
   less than the tidal radius (Table~\ref{tab_sc_data}), and those stars, which
   are identified to be tidal tail members, that is,
   have a distance to the cluster centre
   larger than the tidal radius. Cluster members are marked by black dots,
   members of the leading tidal arm by green dots, members of the trailing
   tidal arm by red dots.
   The positive $x$-axis points towards
   the Galactic centre, the positive $y$-axis towards the Galactic rotation.
   The black arrow indicates the actual direction of motion of the Hyades
   cluster with respect to the Galactic rest frame. It can be clearly seen that
   the leading arm is more populated than the trailing arm. Furthermore,
   the leading tidal arm has also a higher surface density of tidal tail
   members as can be seen in Fig.~\ref{fig_hyades_density}.

   In the upper right panel of Fig.~\ref{fig_hyades} the orientation angle
   and the distance of all stars are shown  using the method described in
   Sect.~\ref{sec_meassure}.
   Stars with an orientation angle smaller
   than 90$^\circ$ belong formally to the leading arm, stars
   with an angle larger than 90$^\circ$ to the trailing arm.
   Within the tidal radius the cluster is well represented by a Plummer
   model \citep{roeser2011a}. Thus,
   in the $\varphi$-$r$-diagram the region between 0 and 180$^\circ$
   is fully populated. The region with
   distances between the tidal radius of 9~pc and $\approx$\,50~pc  
   is still uniformly populated but much sparser than within the bound
   cluster. This corresponds to a relatively spherically shaped
   region around the Hyades cluster, which is sometimes called
   a stellar corona of a star cluster.
   At distances larger than  $\approx$~50\,pc two distinct, sharply confined
   tidal arms are visible.

   The lower left panel of Fig.~\ref{fig_hyades} shows the cumulative
   number of stars in both tidal arms separately as a function of
   the distance to the cluster centre. Up to $\approx$20-30\,pc
   both distributions are nearly equal. Beyond $\approx$30\,pc
   the distributions start to deviate from each other. Up to
   $\approx$100\,pc the cumulative distributions increase
   nearly constantly. The total number of stars in the leading arm
   increases faster than the number of stars in the trailing arm.
   At $\approx$100\,pc both distributions flatten and have again
   a constant but shallower slope. 

   The lower right panel in Fig.~\ref{fig_hyades} shows the cumulative
   normalised asymmetry, $\epsilon(\le r)$, calculated
   by Eq.~(\ref{eq_norm_asymmetry}).

   Figure~\ref{fig_praesepe} shows the same data analysis for the Praesepe star cluster
   (data are taken from \citealt{jerabkova2022a}).
   It can be seen in the lower left panel that  the radial cumulative
   distance distribution at small distances to the star cluster centre shows the same
   functional behaviour as in the case of the Hyades. Within the tidal radius the
   cumulative distributions  rise rapidly and become abruptly shallower in slope at the tidal radius.
   However, the cumulative distributions of both arms are identical up to a distance of
   $\approx$50--70\,pc. Beyond the stellar corona the distributions diverge continuously from each other,
   whereas the trailing arm contains more members than the leading arm, contrary to the Hyades.
   The surface density can be seen in Fig.~\ref{fig_praesepe_density}.

   Figure~\ref{fig_coma_berenices} shows the analysis of the Coma Berenices star cluster
   (data are taken from \citealt{jerabkova2022a}). The radial cumulative distributions have the same
   qualitative trend as those of the Hyades. Within the star cluster and the stellar corona
   both distribution functions are identical. Again, beyond a distance of $\approx$50--70\,pc to the
   cluster centre the distribution diverge continuously from each other. In this case the leading arm
   contains more members. The surface density can be seen in Fig.~\ref{fig_coma_berenices_density}.
   
   Figure~\ref{fig_ngc752} shows the analysis of the NGC~752 star cluster
   (data are taken from \citealt{boffin2022a}). Due to the increasing errors of the
   parallaxes with increasing distance of the star cluster
   the analysis of the tidal tails in three
   dimensions is effected by a distortion of the sample in the $x$-$y$-plane
   \citep{boffin2022a}. Therefore, the stellar sample is analysed in projection
   on the sky.
   
   \section{Monte Carlo simulations}\label{sec_monte_carlo}
   In order to compare the observed asymmetries of tidal tails
   with the degree of asymmetry due to the stochastic
   evaporation of stars from star clusters embedded in a Galactic tidal field,
   a large number of test particle integrations are performed in the Galactic
   gravitational potential.

   \subsection{Numerical model}
   In the simulations a star cluster is set up as a Plummer phase-space
   distribution
   \citep{plummer1911a,aarseth1974a} with initial parameters $b_\mathrm{Pl}$ being the 
   Plummer radius and the total mass, $M_\mathrm{Pl}$.
   The centre of mass of the model, $\vec{r}_\mathrm{Pl}$,
   moves in a Galactic potential as given in \citet{allen1991a}. The orbit
   of each stellar test particle
   is integrated in both gravitational fields, the Plummer potential as the gravitational
   proxy of the star cluster and the full Galactic gravitational potential.
   Thus, the equations of motion of the star $i$ and of the cluster model
   are 
   \begin{equation}
     \vec{a}_i = -\nabla \Phi_\mathrm{Pl} -\nabla \Phi_\mathrm{MW}\,,
   \end{equation}
   \begin{equation}
     \vec{a}_\mathrm{Pl} = -\nabla \Phi_\mathrm{MW}\,.
   \end{equation}
   The test particles are distributed in the model cluster potential
   according to the Plummer phase-space distribution function and are
   integrated in time in the static Galactic potential and a static Plummer
   potential, whose origin is integrated as a test particle in the Galactic
   potential. In a self-gravitating system those particles evaporate from the
   cluster which gained sufficient energy by energy redistribution between
   the gravitationally interacting particles to exceed the binding energy
   to the cluster. In the model cluster all particles are treated as test
   particles and are integrated in a static potential without gravitational
   interaction between the particles. As energy redistribution does not
   occur in this model the evaporation process is simulated as follows:
   If the Plummer sphere were set up in isolation all particles had
   negative energy and were bound to the cluster. As the model cluster
   is positioned in the Galactic external potential the combined potential
   is lowered in two opposite points on the intersecting line defined by the
   Galactic origin and the cluster centre, being the Lagrange points.
   As a consequence, a fraction of the initial set of stars have positive
   energy with respect to the Lagrange points and lead to a continuous
   stream of escaping stars across the clusters's tidal threshold (or pr\'ah
   according to \citealt{kroupa2022a}), with individual escape time scales up to
   a Hubble time. This method has been successfully tested in \citet{fukushige2000a}.
   
   The aim in this work is to quantify the expected
   distribution of the asymmetry between both tidal tails for the observed
   total number of tidal tail members
   due to stochastic evaporation through the Lagrange points. As not all test particles
   escape from the star cluster into the tidal arms a few test runs are required
   to calibrate the total number of test particles, such that the number of
   escaped stars is nearly equal to the observed number of tidal tail members.
  
   As the stellar test particles are performing many revolutions
   around the star cluster centre before evaporating into
   the Galactic tidal field, the equations of motion are integrated
   with a time-symmetric Hermite method \citep{kokubo1998a}.
   The expressions of the accelerations and the corresponding time
   derivatives are listed in App.~\ref{app_hermite} for completeness.
   
   As the total mass of the Plummer model is constant during the simulation,
   models with minimum and maximum star cluster mass are calculated in order
   to enclose the mass loss of real star clusters.
   \cite{roeser2011a} estimated an initial stellar mass of the Hyades of 1100~$M_\sun$
   and determined a current stellar mass gravitationally bound within the tidal radius
   of 275~$M_\sun$.
     Therefore, in the minimum model the mass of the Plummer sphere is given by the current
   stellar mass of the star cluster (Table~\ref{tab_sc_data}), in the maximum model
   the mass of the Plummer sphere is set to four times the current
   stellar mass. The Plummer parameter, $b_\mathrm{Pl}$, is identical in both models.
   
   Furthermore, these two models also take into account
   the possible range of tidal radii:
   a minimum model (current mass) with the smallest tidal radius and a maximum model
   (here 4x the current mass) with a plausible maximum tidal radius.
   
   The orbits of the four star clusters  are integrated
   backwards in time over their assumed life time from their current
   position (Table~\ref{tab_sc_data}) to their location of formation.
   At this position 1000 randomly created Plummer models
   are set up for each maximum and minimum cluster.
   Each configuration is integrated
   forward in time over the assumed age of the respective star cluster.
   Finally all particles outside the tidal radius are assigned to
   their corresponding tidal arm using the method described in
   Sect.~\ref{sec_meassure} and the normalised asymmetry, $\epsilon$,
   is calculated.
   
   \subsection{Statistical asymmetry}
   
\begin{figure}
     \centering
     \includegraphics[width=\columnwidth]{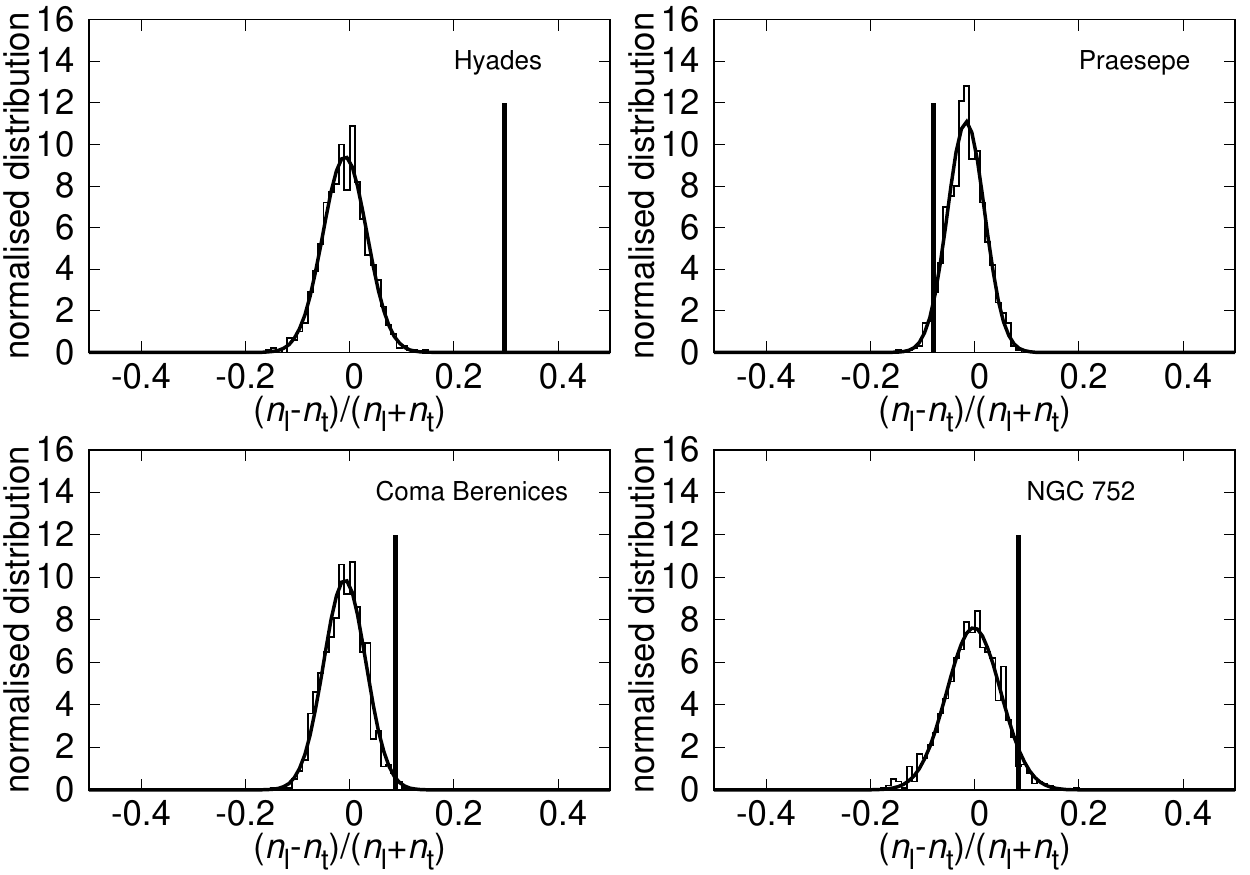}
     \caption{Monte Carlo results for the minimum model: Shown is
       the normalised distribution of the asymmetry in the Monte Carlo
       simulations for each star cluster model.
       The solid curve is the best-fitting Gaussian
       distribution function. The vertical solid line marks the
       observed value.
     }
     \label{fig_monte_carlo_min}
   \end{figure}

\begin{figure}
     \centering
     \includegraphics[width=\columnwidth]{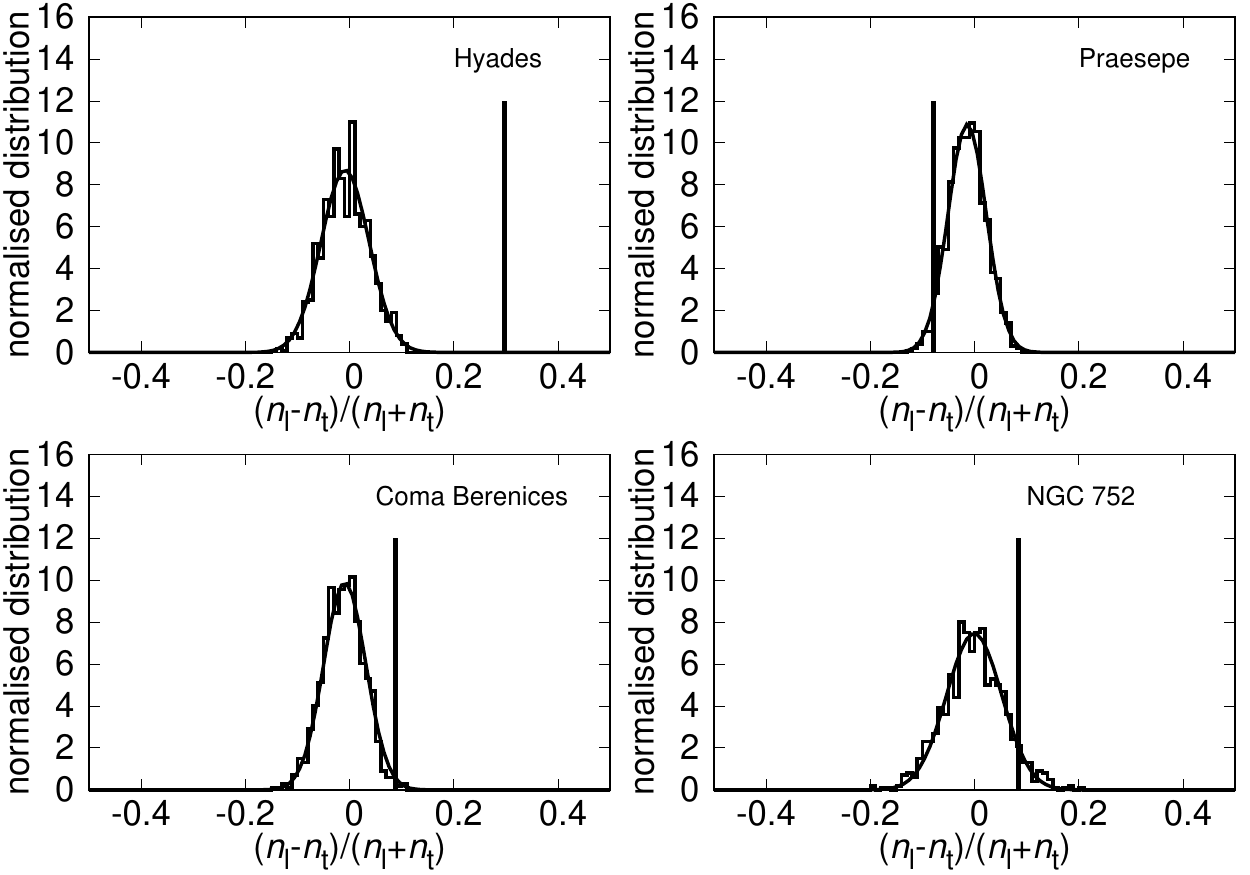}
     \caption{Similar to analysis as in Fig.~\ref{fig_monte_carlo_min}
       but for the maximum model.}
     \label{fig_monte_carlo_max}
   \end{figure}

   The numerically obtained
   distribution of the normalised asymmetry of all four star clusters is shown
   by the histograms in Fig.~\ref{fig_monte_carlo_min} for the minimum models
   and in Fig.~\ref{fig_monte_carlo_max} for the maximum models. The solid curves
   show the best-fitting Gaussian function. The vertical line marks the observed
   asymmetry (Table~\ref{tab_sc_data}). The resulting 
   mean asymmetry, $\mu$, and the dispersion, $\sigma$, of the Gaussian fits
   are listed in Table~\ref{tab_monte_carlo}. 
   In the last column in Table~\ref{tab_monte_carlo} the observed asymmetry is
   tabulated in units of the respective model dispersion. For example, in the case
   of the maximum model the observed asymmetry of the Praesepe cluster is a
   1.7~$\sigma$ event.

   Dispersion and mean value of the minimum and maximum are in good agreement.
   It can be concluded that the mass of the Plummer model has only a small
   influence. In the case of the Praesepe, the Coma Berenices and the NGC~752
   star cluster the observed asymmetries have a  probability less
   than 3~$\sigma$ and can be interpreted as pure statistical events.
   But, if the observed asymmetric tidal tails of the Hyades
   were solely the result of stochastic evaporation, then the asymmetry
   would be at least a 6.7~$\sigma$ event.

   \begin{table}
     \begin{tabular}{cccc}
       Cluster & $\mu$ & $\sigma$ & $|x_\mathrm{obs}-\mu|$\\
       \hline\\
       Hyades (min)   &  -0.0088  & 0.042  & 7.3\,$\sigma$\\
       Hyades (max)   &  -0.0086  & 0.046  & 6.7\,$\sigma$\\
       Hyades (theo.)  & 0 & 0.043 &6.9\,$\sigma$\\
       \hline
       Praesepe (min) &  -0.0163  & 0.036  & -1.7\,$\sigma$\\
       Praesepe (max) &  -0.0142  & 0.037  & -1.7\,$\sigma$\\
       Praesepe (theo.) &  0  & 0.035  & -2.2\,$\sigma$\\
       \hline
       Coma Berenices (min) & -0.0091 &  0.041 & 2.4\,$\sigma$ \\
       Coma Berenices (max) & -0.0092 &  0.040 & 2.4\,$\sigma$ \\
       Coma Berenices (theo.) &  0  & 0.040  & 2.2\,$\sigma$\\
       \hline
       NGC~752 (min) &  -0.0052 &  0.055  &  1.6\,$\sigma$\\
       NGC~752 (max) &  -0.0046 &  0.056 & 1.6\,$\sigma$\\
       NGC~752 (theo.) &  0  & 0.058  & 1.6\,$\sigma$\\
     \end{tabular}
     \caption{\label{tab_monte_carlo}Parameter of Gaussian fits of the
       asymmetry distribution of the different Monte Carlo models.
     }
   \end{table}

   \section{Theoretical considerations}\label{sec_theo}
   In this section a theoretical stochastic evaporation model is developed and
   compared with the
   results of the numerical models of Sect.~\ref{sec_monte_carlo}. 
   \subsection{Theoretical distribution function, $f(\epsilon)$, of the asymmetry}
   The evaporation of a star into one of the tidal tails can be treated as a Bernoulli-experiment.
   Let $p_\mathrm{l}$ be the probability for a star to end up in the leading tail, $(1-p_\mathrm{l})$
   the probability to end up in the trailing tail. For a total number $n$ of tidal tail members
   the probability that $n_\mathrm{l}$ stars are located in the leading arm is given by a
   binomial distribution
   \begin{equation}
     b_n(n_\mathrm{l}) = \binom{n}{n_\mathrm{l}} p_\mathrm{l}^{n_\mathrm{l}} (1-p_\mathrm{l})^{n-n_\mathrm{l}}\,,
    \end{equation}
   with an expectation value of $\mathrm{E}(n_\mathrm{l}) = n p_\mathrm{l}$ and a variance
   $\mathrm{Var}(n_\mathrm{l})=np_\mathrm{l}(1-p_\mathrm{l})$.

   According to the theorem by de Moivre-Laplace, the binomial distribution converges against the
   Gaussian distribution,
   \begin{equation}
     g(n_\mathrm{l}) = \frac{1}{\sqrt{2\pi\sigma_\mathrm{l}^2}} e^{-\frac{(n_\mathrm{l}-\mu_\mathrm{l})^2}{2\sigma_\mathrm{l}^2}}\,,
   \end{equation}
    for increasing $n_\mathrm{l}$
   with an expectation value $\mathrm{E}(n_\mathrm{l})=\mu_\mathrm{l}=np_\mathrm{l}$
   and a variance $\mathrm{Var}(n_\mathrm{l})=\sigma_\mathrm{l}^2=np_\mathrm{l}(1-p_\mathrm{l})$.

   The normalised asymmetry, $\epsilon$, is
   related to the number of members of the leading
   tail, $n_\mathrm{l}$, by 
   \begin{equation}\label{eq_epsilon_n_l}
     \epsilon = \frac{n_\mathrm{l}-n_\mathrm{t}}{n} = \frac{2n_\mathrm{l}}{n}-1\,.
   \end{equation}
   The relation between the distribution function of the
   normalised asymmetry, $f(\epsilon)$, 
   and the distribution of the member number of stars in the leading tail, 
   $g(n_\mathrm{l})$, is given by
   \begin{equation}
     f(\epsilon)\;d\epsilon = g(n_\mathrm{l})\;dn_\mathrm{l}\,.
   \end{equation}
   The distribution function of the asymmetry can then be calculated by
   \begin{equation}
     f(\epsilon) = g(n_\mathrm{l}(\epsilon))\left|\frac{dn_\mathrm{l}}{d\epsilon}(\epsilon)\right|
     = \frac{1}{\sqrt{2\pi\sigma_\epsilon^2}}
     e^{-\frac{\left(\epsilon-\mu_\epsilon\right)^2}{2\sigma_\mathrm{\epsilon}^2}}\,,
   \end{equation}
   with $dn_\mathrm{l}/d\epsilon$ following from Eq.~(\ref{eq_epsilon_n_l}) and
   \begin{equation}
     \sigma_\epsilon^2 = \frac{4\sigma_\mathrm{l}^2}{n^2} =  \frac{4 p_\mathrm{l}(1-p_\mathrm{l})}{n}\,,
   \end{equation}
   and
   \begin{equation}\label{eq_mu_p}
     \mu_\epsilon = \frac{2\mu_\mathrm{l}}{n}-1=2p_\mathrm{l}-1\,.
   \end{equation}

   \subsection{Symmetric evaporation}
   In the case of a symmetric population of the tidal tails, the evaporation
   probabilities into both arms
   are identical, $p_\mathrm{l} = p_\mathrm{t} = \frac{1}{2}$.
   The expectation value and the variance are
   \begin{equation}
     \mu_\epsilon = 0\;\;\;,\;\;\;\sigma_\epsilon=\frac{1}{\sqrt{n}}\,.
   \end{equation}
   For all four clusters the theoretically expected asymmetry due to stochastic
   evaporation, if the tidal tails are symmetrically populated,
   is listed in Table~\ref{tab_monte_carlo}. It can be seen that in all four cases
   the theoretical values are close to the numerically obtained ones. 

   \subsection{Asymmetric evaporation}\label{sec_asy_evap}
   Now, consider the case that the evaporation and the distribution processes
   within the vicinity of the star
   cluster into the tidal arms were asymmetric.
   According to Eq.~\ref{eq_mu_p} the expectation value of the
   asymmetry, $\mu_\epsilon$,
   increases, that is, more stars evaporate into  the leading arm than into the trailing arm, if the population
   probability, $p_\mathrm{l}$, of the leading arm increases.

\begin{figure}
     \centering
     \includegraphics[width=\columnwidth]{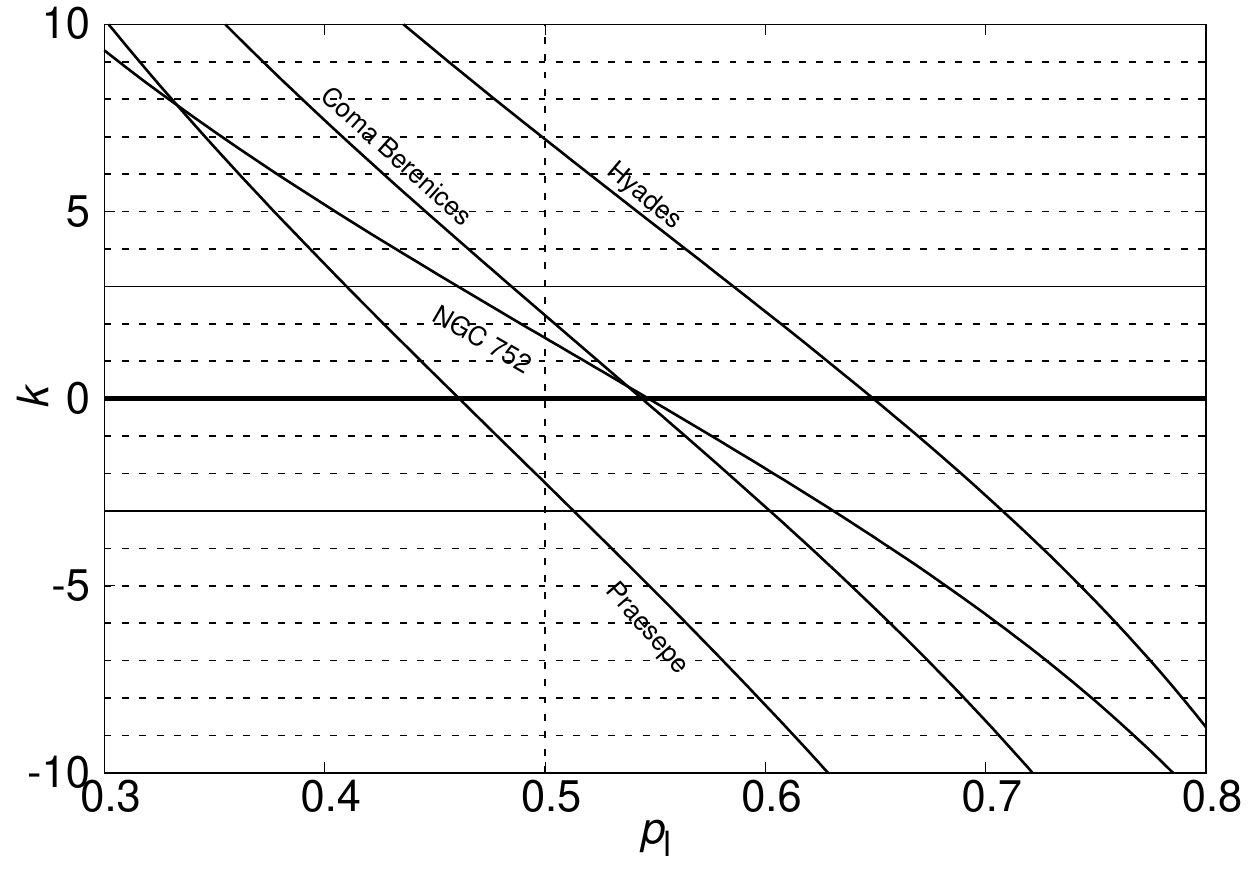}
     \caption{Shown is the probability of the observed asymmetry
       of all four star clusters
       as a multiple of the dispersion
       in dependence of the assumed evaporation probability
       into the leading arm, $p_\mathrm{l}$. See Sect.~\ref{sec_asy_evap} for details.}
     \label{fig_sigma_p}
   \end{figure}

   For a given observed asymmetry, $\epsilon_\mathrm{obs}$, the probability of this event can
   be calculated as multiples, $k$,
   of the dispersion for the assumed evaporation probability $p_\mathrm{l}$, 
   \begin{equation}\label{eq_k}
     k = \frac{\epsilon_\mathrm{obs}-\mu_\epsilon(p_\mathrm{l})}{\sigma_\epsilon(p_\mathrm{l})}
     = \frac{\sqrt{n}(\epsilon_\mathrm{obs}-2p_\mathrm{l}+1)}{2\sqrt{p_\mathrm{l}(1-p_\mathrm{l})}}\;.
   \end{equation}
   Fig.~\ref{fig_sigma_p} shows this function for all four star clusters. The vertical dashed line marks
   the case of symmetrically populated tidal tails. For example, the Hyades have an observed asymmetry of
   $\epsilon_\mathrm{obs} =0.298$ (Table~\ref{tab_sc_data}).
   In the case of  a symmetric
   evaporation, $p_\mathrm{l}=\frac{1}{2}$, the theoretical dispersion is
   $\sigma = 0.043$. Thus, the probabillity of this event is 0.298/0.043 = 6.9\,$\sigma$.
   This point is marked in Fig.~\ref{fig_sigma_p} by the intersection of the solid line labeled
   with Hyades and the vertical dashed line.

   On the other hand, in order to increase the event probability (decreasing $k$)
   the probability of evaporation into the leading arm needs to be increased. For given
   $k$, Eq.~(\ref{eq_k}) can be solved for the required
   evaporation probability, $p_\mathrm{l}$, into the leading arm. The emerging
   quadratic equation leads to two solutions of $p_\mathrm{l}$,
   \begin{equation}
     p_{1,2} = -\frac{a}{2}\pm\sqrt{\left(\frac{a}{2}\right)^2-b}\,,
   \end{equation}
   where
   \begin{equation}
     a=-\frac{k^2+n(\epsilon_\mathrm{obs}+1)}{k^2+n}\;\;\;\mathrm{and}\;\;\;
     b = \frac{n(\epsilon_\mathrm{obs}+1)^2}{4k^2+4n}\,.
   \end{equation}
   If the observed asymmetry of the Hyades should be
   a 3\,$\sigma$ event then an evaporation probability into the leading arm of approximately
   $p_\mathrm{l} = 0.585$ is required.

   \section{Discussion and Conclusions}
   The common assumption that both tidal tails of star clusters, moving on nearly
   circular obits around the Galactic centre, evolve equally has recently faced a challenge
   as the analysis of Gaia data reveal asymmetries in the tidal tails of four
   nearby open star clusters.

   Because the evaporation of stars can be treated as a
   stochastic process the normalised difference of the number of member stars of both tails
   should follow a distribution function. This distribution has been quantified here
   by use of Monte Carlo simulations of test particle configurations integrated in the full
   Galactic potential and compared with a theoretical approach. It emerges that the theoretical
   and numerical results agree with each other.
   
   Comparing the individual distribution functions of the asymmetry with
   the observed ones, it
   can be concluded that the observed
   asymmetry of Praesepe, Coma Berenices and NGC~752 might be the result
   of the stochastic nature of the evaporation of stars through both Lagrange points.
   On the other hand, the asymmetry of the Hyades is a 6.7~$\sigma$ event.
   In order to interpret the asymmetry as a 3~$\sigma$ event,
   asymmetric evaporation probabilities into the leading arm of  58.5\%
   and into the trailing arm of 41.5\% are required.
   
   It might be speculated that external effects
   might lead to an additional broadening
   of the distribution function of the asymmetry.
   Assuming a different value for the Galactic rotational velocity or a
   different position
   of the Sun than used in this work might not lead to a larger scatter as the
   50/50\% evaporation probabilities at both Lagrange points are not expected to
   vary in an almost flat rotation curve.
   
   A stronger effect on the asymmetry of the tidal tails might 
   be due to local deviations from a logarithmic
   potential (as required in the case of a flat rotation curve). Such local variations
   can be a result of an interaction with
   a Galactic bar or spiral arms \citep[cf.][]{bonaca2020a,pearson2017a}. How strong this effect will be
   can be hardly estimated and will be explored in further numerical studies.
   However, the main result here, that the pure evaporation of stars through
   the Lagrange points is basically a simple
   Bernoulli process, remains solid.
   
   If the asymmetric evaporation probabilities have an internal origin, then larger
   asymmetries in the star cluster potential and the kinematics of the evaporation
   process is required 
   than Newtonian dynamics can provide \citep{kroupa2022a}.
   On the other hand the asymmetry might be due to an external perturbation,
   for example
   through the encounter with a molecular cloud \citep{jerabkova2021a}.
   However, the detailed analysis of the asymmetry in the tidal tails of
   the Hyades reveals
   that the process leading to the asymmetry must affect both arms equally in terms
   of the qualitative structure. The bending of the radial number distributions occur at the same
   distance to the star cluster but with different strength (Fig.~\ref{fig_hyades}, lower left panel).
   If an encounter with an external
   object had occurred on one side of the star cluster then it is expected that more than only the total number
   of tidal tail members is reduced. Instead, the functional form of the
   cumulative number distribution in both arms would be completely different
   and rapid changes of the slopes of the cumulative number distribution
   would not be expected to occur at the same distance from the star cluster
   centre in both tidal tails on opposite sites of the star cluster
   as is observed in the tidal tails of the Hyades (Fig.~\ref{fig_hyades}, lower left panel).
   
%   \begin{acknowledgements}
%   \end{acknowledgements}

   \bibliography{n-body,star-cluster,star-formation,milkyway,cmf,onc,OB-star,mond,submitted,prep}
   \bibliographystyle{aa}
   
   \begin{appendix}
     \section{Hermite formulae for the Allen-Santillan  potential}
     The total Galactic gravitational field has contributions from
     three components: Galactic bulge, disk and halo. The rotationally
     symmetric gravitational potentials are taken from \citet{allen1991a}.
     The time-symmetric Hermite integrator from \citet{kokubo1998a}
     requires the accelerations $\vec{a}$ and their
     time derivatives $\vec{j}=\vec{\dot{a}}$ and are listed below.
     The following notations are used:
     $\vec{r} = (x,y,z)$, $\boldsymbol{\rho} = (x,y,0)$,
     $\vec{z} = (0,0,z)$,
     $\vec{\dot{r}} = \vec{v}=(\dot{x},\dot{y},\dot{z})$,
       $\boldsymbol{\dot{\rho}} = (\dot{x},\dot{y},0)$,
     $\vec{\dot{z}} = (0,0,\dot{z})$. 
    
     \label{app_hermite}
     \begin{equation}
       \Phi(\rho,z) =
       \Phi_\mathrm{bulge}(\rho,z)
       +\Phi_\mathrm{disk}(\rho,z)
       +\Phi_\mathrm{halo}(\rho,z)
     \end{equation}
     \begin{equation}
       \vec{a} =
       \vec{a}_\mathrm{bulge}
       +\vec{a}_\mathrm{disk}
       +\vec{a}_\mathrm{halo}
       =-\nabla\Phi_\mathrm{bulge}
       -\nabla\Phi_\mathrm{disk}
       -\nabla\Phi_\mathrm{halo}
     \end{equation}
     \begin{equation}
       \vec{j} =
       \vec{j}_\mathrm{bulge}
       +\vec{j}_\mathrm{disk}
       +\vec{j}_\mathrm{halo}
     \end{equation}

     \subsection{bulge}
     \begin{equation}
     \Phi_\mathrm{bulge}(\rho,z)
     =-GM_1\frac{1}{\sqrt{\rho^2+z^2+b_1^2}}
     \end{equation}
     \begin{equation}
     \vec{a}_\mathrm{bulge}
     =-GM_1\left(\rho^2+z^2+b_1^2\right)^{-\frac{3}{2}}\vec{r}
     \end{equation}
     \begin{equation}
     \vec{j}_\mathrm{bulge}
     =-GM_1\left(\rho^2+z^2+b_1^2\right)^{-\frac{3}{2}}
     \left(\vec{v}-3\frac{{\vec{\dot{r}}}\bullet\vec{r}}{\rho^2+z^2+b_1^2}\vec{r}\right)
     \end{equation}

     \subsection{disk}
     \begin{equation}
     \Phi_\mathrm{disk}(\rho,z)
     =-GM_2\frac{1}{\sqrt{\rho^2+\left( a_2 + \sqrt{z^2+b_2^2}\right)^2}    }
     \end{equation}
     \begin{eqnarray}
       \vec{a}=-GM_2\left(\rho^2+
       \left(a_2+\left(z^2+b_2^2\right)^{\frac{1}{2}}\right)^2
       \right)^{-\frac{3}{2}}\\
       \left(\boldsymbol{\rho}+\left(1+\frac{a_2}{\sqrt{z^2+b_2^2}}\right)\vec{z}\right)
     \end{eqnarray}
     \begin{eqnarray}
       \vec{j}=3M_2 G\left(\rho^2+\left(a_2+\left(z^2+b_2^2\right)^{\frac{1}{2}}\right)^2\right)^{-\frac{5}{2}}\cdot\hfill\hfill\\
       \left(
       \boldsymbol{\rho}\bullet\dot{\boldsymbol{\rho}}+
       \left(1+\frac{a_2}{\sqrt{z^2+b_2^2}}\right)\vec{z}\bullet\dot{\vec{z}}\right)
       \left(\boldsymbol{\rho}+\left(1+\frac{a_2}{\sqrt{z^2+b_2^2}}\right)\vec{z}\right)\\
       -M_2 G\left(\rho^2+\left(a_2+\left(z^2+b_2^2\right)^{\frac{1}{2}}\right)^2\right)^{-\frac{3}{2}}\cdot\\
       \left(\dot{\boldsymbol{\rho}}-\frac{a_2\vec{z}\bullet\dot{\vec{z}}}{\left(z^2+b_2^2\right)^{\frac{3}{2}}}\vec{z}
       +\left(1+\frac{a_2}{\sqrt{z^2+b_2^2}}\right)\dot{\vec{z}}\right)
     \end{eqnarray}
     
     \subsection{halo}
     \begin{equation}
       \Phi_\mathrm{halo}=-\frac{GM_\mathrm{halo}(\le r)}{r} - \frac{GM_3}{(\gamma-1)a_3}
       \left[-\frac{\gamma-1}{1+\left(\frac{r}{a_3}\right)^{\gamma-1}}
       +\ln\left(1+\left(\frac{r}{a_3}\right)^{\gamma-1}\right)\right]_{r}^{100}\,,
     \end{equation}
     where
     \begin{equation}
       M(\le r) = \frac{M_3\left(\frac{r}{a_3}\right)^\gamma}{1+\left(\frac{r}{a_3}\right)^{\gamma-1}} 
     \end{equation}
     is the total halo mass within the Galactocentric radius $r$.
     \begin{equation}
     \vec{a}_\mathrm{halo}
     =-\frac{GM(\le r)}{r^3}\vec{r}
     \end{equation}
     \begin{eqnarray}
     \vec{j}_\mathrm{halo}
     =-\frac{G M(r)}{r^3}
     \left(\vec{v}-\frac{\vec{r}\bullet\vec{v}}{r^2}\left(\gamma-3-\frac{(\gamma-1)\left(\frac{r}{a_3}\right)^{\gamma-1}}{1+\left(\frac{r}{a_3}\right)^{\gamma-1}}\right)\vec{r}\right)
     \end{eqnarray}
     \subsection{Units and parameters}
     In \citet{allen1991a} the spatial variables of the potential have
     the unit kpc, the velocity the unit 10\,km\,s$^{-1}$
     and the potential the unit 100\,km$^2$\,s$^{-2}$.
     Thus, the accelerations
     needs to be multiplied by 0.104572 in order to have the unit
     pc\,Myr$^{-2}$ and their time derivatives, $\vec{j}$,
     by $1.06936\times 10^{-3}$ to have the unit pc\,Myr$^{-3}$.
     The mass has the unit $2.3262\times 10^{7}\,M_\sun$
     and is scaled such that the gravitational constant is $G=1$.
     The parameters of the potentials are summarised in
     Table~\ref{tab_pot_value}.
     
      \begin{table}
        \begin{tabular}{cc}
          component & parameter \\\hline
          bulge & $M_1$ = 606.0\\
          bulge & $b_1$ = 0.3873\\
          disk & $M_2$ = 3690.0\\
          disk &$a_2$ = 5.3178\\
          disk &$b_2$ = 0.2500\\
          halo &$M_3$ = 4615.0\\
          halo &$a_3$ = 12.0\\
          halo &$\gamma$ = 2.02\\
        \end{tabular}
        \caption{\label{tab_pot_value} Parameters of the potential
          used in this work.}
      \end{table}

      \subsection{Hermite formulae for the Plummer sphere}
      \begin{equation}
      \Phi_\mathrm{Pl}(r)
     =-GM_\mathrm{Pl}\frac{1}{\sqrt{r^2+b_\mathrm{Pl}^2}}
     \end{equation}
     \begin{equation}
     \vec{a}_\mathrm{Pl}
     =-GM_\mathrm{Pl}\left(r^2+b_\mathrm{Pl}^2\right)^{-\frac{3}{2}}\vec{r}
     \end{equation}
     \begin{equation}
     \vec{j}_\mathrm{Pl}
     =-GM_\mathrm{Pl}\left(r^2+b_\mathrm{Pl}^2\right)^{-\frac{3}{2}}
     \left(\vec{v}-3\frac{{\vec{\dot{r}}}\bullet\vec{r}}{r^2+b_\mathrm{Pl}^2}\vec{r}\right)
     \end{equation}
  
   \end{appendix}

\end{document}